*Article*

# Total electron content PCA-NN model for middle latitudes

Anna Morozova [1,2,*], Teresa Barata [1,3], Tatiana Barlyaeva [1] and Ricardo Gafeira [1,2]

[1] Instituto de Astrofísica e Ciências do Espaço, University of Coimbra, OGAUC, Coimbra, Portugal
[2] Physics Department, FCTUC, University of Coimbra, OGAUC, Coimbra, Portugal
[3] Earth Sciences Department, FCTUC, University of Coimbra, OGAUC, Coimbra, Portugal
* Correspondence: AM: annamorozovauc@gmail.com, anna.morozova@uc.pt

**Abstract:** A regression-based model was previously developed to forecast the total electron content (TEC) at middle latitudes. We present a more sophisticated model using neural networks (NN) instead of linear regression. This regional model prototype simulates and forecasts TEC variations in relation to space weather conditions. The development of a prototype consisted of the selection of the best set of predictors, NN architecture and the length of the input series. Tests made using the data from December 2014 to June 2018 show that the PCA-NN model based on a simple feed-forward NN with a very limited number (up to 6) of space weather predictors performs better than the PCA-MRM model that uses up to 27 space weather predictors. The prototype is developed on a TEC series obtained from a GNSS receiver at Lisbon airport and tested on TEC series from three other locations at middle altitudes of the Eastern North Atlantic. Conclusions on the dependence of the forecast quality on longitude and latitude are made.

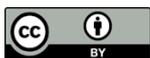

## 1. Introduction

Forecasting of ionosphere parameters like the total electron content (TEC) is essential for the improvement of the quality of the GNSS-based services, including operations of the unmanned air-born, floating, and land vehicles for surveys and investigations of hard-to-reach objects and locations, navigation of newly developed and autonomous vehicle to transport goods and passengers, GNSS-assisted landing procedures for commercial aviation, GNSS positioning during rescue operations [1]. One of the actively developed types of models is empirical (based on the observational data and statistical analysis of such data) models that relate observed variations of, e.g., TEC with such external forcings as solar flares, geomagnetic storms, and other space weather events. Some of these models are global, but others are regional or made to forecast TEC, even for a single location.

Neural networks (NN) [see, e.g., 3-4] and NN in combination with different decomposition methods as, for example, the ensemble empirical mode decomposition (EEMD) [5] or the principal component analysis (PCA) [see 6 and references therein for a review of PCA-NN based TEC models, as well as, e.g., 7-8] of TEC data are often used to forecast TEC. Most of them use a set of space weather parameters (most often used are the solar F10.7 index and the geomagnetic indices Kp and Dst) as predictors for TEC variations. In general [see 6 and references therein], such models allow for forecasting TEC with RMSE or MAE (root mean squared error / mean absolute error, respectively) in a range from ~2 to 15 TECu depending on the season, geomagnetic, and solar activity level during the studied time interval, geographic location and the type of TEC data (single-station data or global ionospheric maps).

In this work, we present a PCA-NN model prototype that combines PCA and NN to forecast TEC for a single station at middle latitudes. The data used to build and validate the model are described in Section 2. The prototype development (model description, comparison to a previous PCA-based model developed in our group) and metrics used to estimate the forecasting quality are presented in Section 3. Section 4 presents results obtained when the developed prototype is applied to different TEC data sets. Section 5 contains conclusions.

**2. Data**

*2.1 TEC data*

The TEC series is obtained in Lisbon (Portugal) using a GNSS receiver with the SCINDA software [see 9, and references therein] that has been active from November 2014 to March 2019 in the Lisbon airport (38.8ºN, 9.1ºW) in the frame of the ESA Small ARTES Apps project SWAIR (Space Weather and GNSS monitoring services for Air Navigation). The installed equipment was a NovAtel EURO4 receiver with a JAVAD Choke-Ring antenna. The SCINDA software automatically converts the GNSS receiver data to TEC.

The data originally of 1 min time resolution were averaged to obtain the 1 h series. The raw TEC data were processed using a "SCINDA-Iono" toolbox for MATLAB and scripts for R [9-10] developed by our group. The data between December 2014 and February 2019 are available at [11] and [12]. The first part of this dataset which includes only the data for 2015, is described in [9], and TEC variations related to the solar flares and geomagnetic disturbances of 2015 were analysed in [13]. An extension of the TEC series (data for December 2014 and from January 2016 to February 2019) is described in [10].

The calibration procedure was not performed during the installation of this receiver. Therefore, we performed a provisional calibration of the TEC records using TEC data from the Royal Observatory of Belgium (ROB) as a reference. The calibration procedure is described in [10]. Since for Lisbon UT = LT, no time conversion was applied. The SCINDA data have several gaps that were filled with ROB TEC data.

The dataset covering the year 2015 was used to build and test the PCA-MRM model to forecast TEC [6]. This model is used further as a reference model to test the performance of the newly developed prototype for the PCA-NN model.

After the prototype was developed, it was tested on TEC data obtained from a different source: The TEC series were collected for three locations: Continental Portugal, the area around Lisbon; the Azores archipelago, area around the Santa Maria; and São Miguel islands and the Madeira archipelago, area around Funchal city—see Figure 1. This dataset (hereafter, RENEP dataset) was obtained from RINEX data files provided by RENEP (Rede Nacional de Estações Permanentes GNSS, https://renep.dgterritorio.gov.pt/ (accessed on 30 May 2023)) geodetic receivers at Cascais (near Lisbon, grey dot in Fig. 1), Furnas (Azores archipelago, São Miguel island, blue dot in Fig. 1) and Funchal (Madeira archipelago, Madeira island, green dot in Fig. 1). The coordinates of all these receivers are shown in Table 1.

**Table 1.** Coordinates of GNSS receivers.

| Location | Latitude | Longitude |
|---|---|---|
| SCINDA (Lisbon) | 38.7° N | 9.14° W |
| Cascais | 38.7° N | 9.4° W |
| Furnas (Azores) | 37.8° N | 25.3° W |
| Funchal (Madeira) | 32.7° N | 16.9° W |

RENEP provides data in the RINEX 2.11 format. These files were processed and calibrated using the GNSS Lab (http://www.gnss-lab.org/ (accessed on 26 May 2023))—see [14-16] and TEQC (https://www.unavco.org/software/data-processing/teqc/teqc.html (accessed on 26 May 2023, see the website for the updated information about the manufacturer and currently available versions) software. TEC series for all the receivers were averaged to have 1h time resolution. The RENEP data series has a gap from July 2015 to December 2016.

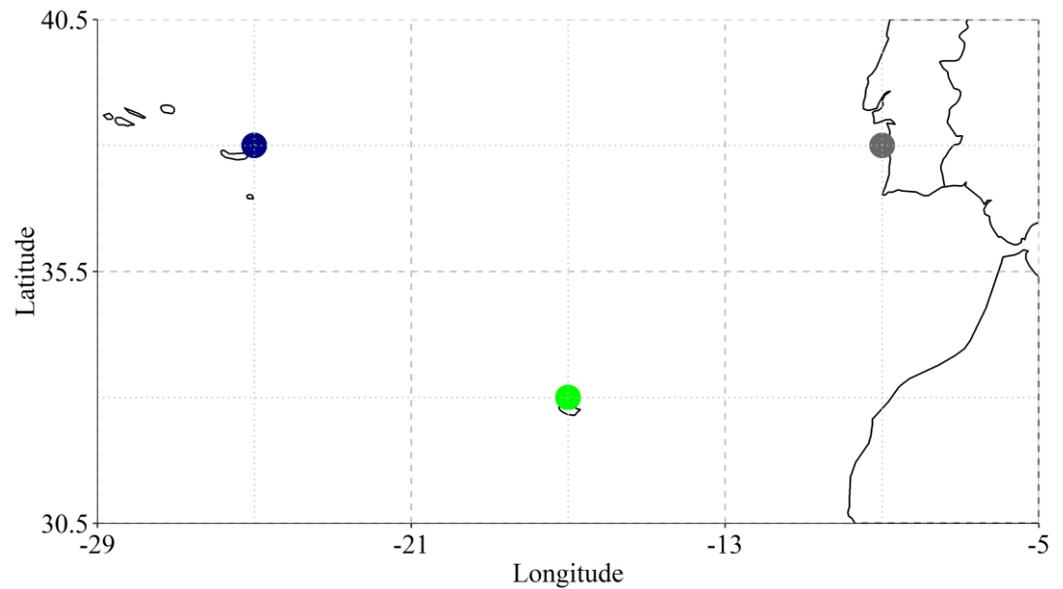

**Figure 1.** Colored circles show the approximate location of GNSS receivers at Continental Portugal (grey) and the Azores (blue) and Madeira (green) archipelagos. The colors of the circles correspond to the colors of the lines in Figure 5.

*2.2 Space weather parameters*

Results of [6, 13, 17] were used to define the set of space weather parameters (SWp) which are predictors for TEC variations in PCA-NN models. In particular, the daily number of flares of different classes (B and C, and M) and the total number of flares (N) are used to understand the most influential (for TEC) class of flares: the total number of flares of any class, the number of the most abundant flares (B and C classes) or the number of moderate flares (M class). Since the number of X flares during the studied period is very small (5 flares in ~2.5 years), the effect of the X flares on the model's performance was not studied.

Initially, three types of SWp were used to forecast TEC variations, both for the PCA-MRM and PCA-NN models (all SWp series have 1d time resolution):

1. Parameters characterizing the interplanetary medium:
1.1. scalar of the interplanetary magnetic field (IMF), B in nT;
1.2. the X, Y and Z components of IMF in the GSM frame, $B_X$, $B_Y$ and $B_Z$, respectively, all in nT;
1.3. the solar wind flow pressure (p in nPa), proton density (n in n/cm3) and plasma speed (v in km/sec).
2. Geomagnetic indices:
2.1. the disturbance-time index Dst;
2.2. the global ap index;
2.3. the daily sums of the Kp index;
2.4. the auroral electrojet index AE characterizing the auroral activity in the polar regions.
3. Parameters characterizing the solar UV and XR fluxes:
3.1. proxies for the solar UV irradiance:
3.1.1. the Mg II composite series[18-19] based on the measurements of the emission core of the Mg II doublet (280 nm), hereafter Mg II;
3.1.2. the F10.7 index that, is a solar radio flux at 10.7 cm (2800 MHz), which was found to be very well correlated with the solar UV flux [18];
3.2. a proxy for the solar XR irradiance: the measurements of the Solar EUV Experiment (SEE) for the NASA TIMED mission at the wavelength 0.5 nm, hereafter XR;
3.3. the daily number of the solar flares

3.3.1. of the B and C classes, hereafter Number of C flares, C or C.f.;

3.3.2. of the class M, hereafter Number of M flares, M or M.f.;

3.3.3. total daily number of the solar flares of any class from B to X, hereafter Number of flares, N, or N.f.;

The data on the solar wind properties were obtained from the OMNI data base. The information about the solar flares observed during the analysed time interval was obtained through the NOAA National Geophysical Data Center (NGDC). Only flares that occurred during the local daytime were considered.

Also, since the solar wind coupling functions [20] are used in some studies to explain or model TEC [21], we tested some of them as predictors for TEC: $d\Phi_{MP}/dt$, $E_{WAV}$, $E_{WV}$, $\epsilon_3$, $E_{KLV}$, $E_{KL}$, $v \cdot B_S$, $E_{SR}$, $E_{TL}$ [see 20 for description].

Following our previous studies [13] and some other works [e.g., 22-23], we used lags of 1 and 2 days between the TEC and SWp series (SWp series lead).

Supplementary Table S1 shows, as an example, the correlation coefficients calculated for the SWp series for 2015 (only $|r| \geq 0.68$ are shown). As one can see, many SWp are strongly cross correlated. The effect of those cross-correlations on the model's forecasting quality in case the correlated SWp are used as predictors together is not clear a priori. Thus, one of the goals of the presented work was to study the effect of the correlated predictor on the forecast skills of the models.

Due to the availability of the TEC data and the AE index at the time of the model development, all the data sets (TEC and SWp) are limited to the time interval from December 2014 to June 2018.

## 3. PCA-NN model prototype development

### 3.1. PCA-based models

The model proposed in this work is a PCA-based model consisting of a combination of the principal component analysis (PCA) and a certain method used to forecast TEC parameters considering SWp as predictors.

The first step of a PCA-based model is the principal component analysis (PCA) applied to the 1h TEC series. The input data set is used in PCA to construct a covariance matrix and calculate corresponding eigenvalues and eigenvectors. The eigenvectors are used to calculate principal components (PC) and their amplitudes, which we call the empirical orthogonal functions (EOF). The eigenvalues allow us to estimate the explained variances of the extracted modes. PCs are orthogonal and conventionally non-dimensional, and EOFs are in TECu. The full descriptions of the PCA method can be found in (e.g.) [24-26].

For the PCA-based model, the PCA input matrices were constructed so that each column contains 24 observations (1 h data) for a specific day. The daily mean TEC values were removed before the series was submitted to PCA. The number of columns, L, is equal to the length of the studied interval in days. Thus, PCA allows us to obtain daily variations of different types as PCs and the amplitudes of those daily variations for each day of the studied time interval as corresponding EOFs. Consequently, the PC series have 1h time resolution (24 values each), and EOF series has a 1d time resolution (L values each). In this work, only the 1st and 2nd PCA modes, PC1 and EOF1, and PC2 and EOF2, respectively, were used to reconstruct TEC variations [6,13].

One of the advantages of the PCA-based models is that there is no need for any assumption on the phase and amplitude or seasonal/regional features of TEC daily variations: the daily variations of correct shapes are extracted automatically by PCA from the input TEC data as PCs (Fig. 2).

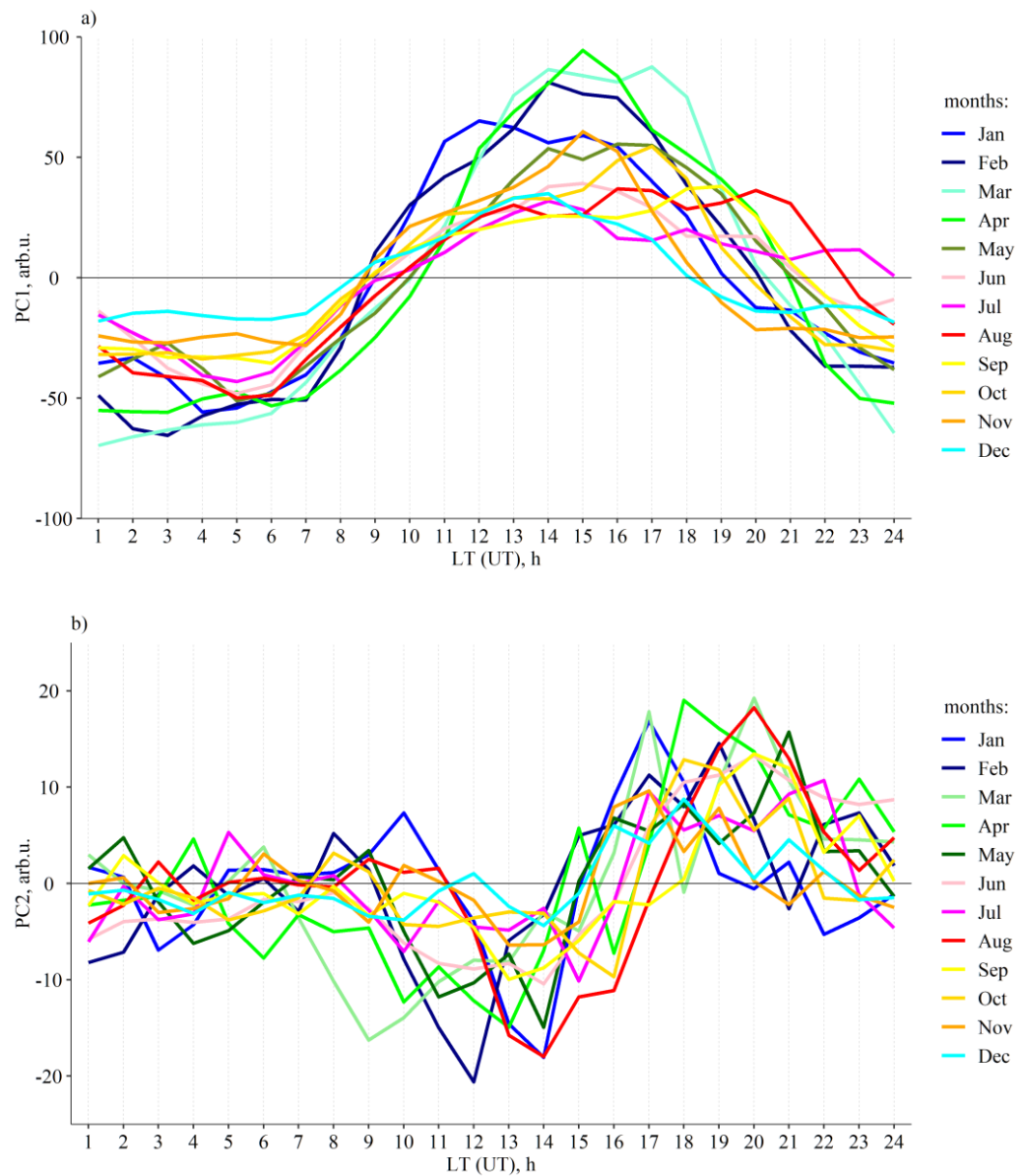

**Figure 2.** Examples of TEC PCs: PC1s (a) and PC2s (b) calculated for months from January to December of 2015.

The second step of PCA-based models consists of reconstruction/forecasting of the *TEC parameters* (i.e., *daily mean TEC* values and the amplitudes of the first two modes, *EOF1* and *EOF2*) using a certain method. Previously, a PCA-MRM model was proposed where the multiple regression models (MRM) were used for the reconstruction/forecasting of TEC parameter – see [6].

Section 3.3 gives a summary of the PCA-MRM performance based on the data of 2015.

In the present work, a new PCA-NN model is proposed (Fig. 3), where the TEC parameters are forecasted using a neural network (NN). A general concept of NN models and the PC-NN models' performance are discussed in Section 3.4.

*3.2 Forecasting skills (metrics)*

The quality of the forecast of the TEC series by the proposed models was characterised by a set of standard metrics or skills described below.

Similarities between series (e.g., between the forecasted and observed TEC) were analysed using the Pearson correlation coefficients, *r*. The significance of the correlation coefficients was estimated using the Monte Carlo approach with artificial series constructed by the "phase randomization procedure" [27]. The obtained statistical significance (*p value*) considers the probability of a random series to have the same or higher absolute value of r as in the case of a tested pair of the original series.

The quality of the forecast was also estimated using the following parameters: the root mean squared error *RMSE* (Eq. 1), the explained variance *ExpV* (Eq. 2), the coefficient of determination $R^2$ (Eq. 3), the mean absolute error *MAE* (Eq. 4), the maximum error *MaxE* (Eq. 5), a percentage of the forecasted series with ΔTEC in certain limits. The forecasting quality of a model is better if it has lower values of MAE and RMSE, higher values of r, $R^2$ and ExpV, and a higher percentage of days with ΔTEC in a certain small range.

$$RMSE = \sqrt{\Sigma(y_i - \hat{y}_i)^2/N} \tag{1}$$
$$ExpV = 1 - \sigma_{\Delta y}^2/\sigma_y^2 \tag{2}$$
$$R^2 = 1 - \Sigma(y_i - \hat{y}_i)^2/\Sigma(y_i - \bar{y})^2 \tag{3}$$
$$MAE = mean\ |\Delta TEC| = \Sigma|y_i - \hat{y}_i|/N \tag{4}$$
$$MaxE = max(|y_i - \hat{y}_i|) \tag{5}$$

where $y_i$ and $\hat{y}_i$ are the observed and modelled series, respectively; $\bar{y}$ and $\sigma_y$ are the mean and SD for $y_i$, respectively; $\sigma_{\Delta y}$ is SD for the $\Delta TEC = (y_i - \hat{y}_i)$ series; and N is the length of the series.

*3.3 PCA-MRM model*

As was said before in Section 3.1, the second step of a PCA-MRM model is a construction of a linear multiple regression model (MRM), i.e., a linear regression model with not one but a number of regressors [6,28]. In these MRMs, the TEC parameters are dependent variables and SWp are regressors. MRMs for a studied time interval of L days for different TEC parameters are built separately and may have different SWp regressors. MRMs were constructed to fit the data with all possible combinations of regressors, and the one with a minimal squared coefficient of multiple determination adjusted for the number of degrees of freedom ($R_{adj}^2$) was selected. Thus, the final model is built using a "best subset" of regressors, ensuring that only those most influential regressors for a particular TEC parameter and for a specific time interval were selected.

A detailed description of the PCA-MRM model and its performance can be found in[6]. Here we give a concise summary of its performance. The best performance was obtained for PCA-MRM models with *L* (length of the input data sets) equal to 31 or 32 days. This length of the input data sets seems to be a compromise: the shorter length may result in a better representation of the TEC daily modes but will not be sufficient for the construction of reliable MRMs with so many regressors; on the other hand, larger *L* will allow constraining the regression coefficients well, but the TEC daily modes may be resolved with lower quality because of the seasonal changes of the TEC daily variation.

The forecasting quality of the PCA-MRM model was studied on the hourly (1h TEC series) and the daily (daily mean and daily maximum TEC series) time scales, during quiet days (no solar flares, no geomagnetic disturbances), days with solar flares, and days with geomagnetic disturbances, during different months, and during different hours of a day. For the test time interval (February to December 2015) and for a mid-latitudinal location (Lisbon, Portugal), the PCA-MRM model allows 90% confidence intervals of 6 TECu for day hours and 3 TECu for night hours – see Table 2 for this and other skills values. These confidence intervals are calculated using all available days and do not consider the solar or geomagnetic activity level. Table 3 shows MAE obtained for the PCA-MRM model for different types of days: quiet days (days without M or X flares and no more than one flare of the class C or below, and without geomagnetic disturbances), days with M or X flares and/or with more than one flares of class C or below, geomagnetically disturbed days without and with flares.

**Table 2.** Forecasting skills of PCA-based models. The metrics for the 1h series are calculated considering all data points ("all"), all data points (hours) but averaged for all individual days ("all (daily means)") or separately for the day/night hours and averaged for all individual days ("day/night (daily means)"). Better forecasting skills between the two models are in bold.

| Metrics | Time resolution | Hours | PCA-MRM (2015 only) | PCA-NN (2014-2018) |
|---|---|---|---|---|
| r (TEC$_{for}$ vs TEC$_{obs}$) | 1h | all | 0.89 | **0.92** |
| | 1d mean | – | 0.88 | **0.94** |
| RMSE, TECu | 1h | all | 4.3 | **2.7** |
| | | all (daily means) | 3.7 | **2.7** |
| | | day (daily means) | 4.5 | **3.3** |
| | | night (daily means) | 2.4 | **2.0** |
| | 1d mean | – | 2.8 | **1.7** |
| MAE, TECu | 1h | all | 2.9 | **1.9** |
| | | all (daily means) | 2.9 | **1.9** |
| | | day (daily means) | 3.9 | **2.3** |
| | | night (daily means) | 2.0 | **1.5** |
| | 1d mean | – | 2.0 | **1.2** |
| MaxE, TECu | 1h | all | 44.7 | **35.3** |
| | 1d mean | – | 15.3 | **13.6** |
| ExpV | 1h | all | 0.78 | **0.85** |
| | 1d mean | – | 0.77 | **0.89** |
| $R^2$ | 1h | all | 0.80 | **0.85** |
| | 1d mean | – | 0.77 | **0.89** |
| 90% confidence level, TECu | 1h | all | ±5.0 | **±4.0** |
| | | day (daily means) | ±6.0 | **±4.8** |
| | | night (daily means) | **±3.0** | ±3.2 |

**Table 3.** MAE (in TECu) of PCA-based models calculated for different types of days in 2015. Better forecasting skills between the two models are in bold. The time resolution of the series is 1h.

| Days | Hours | PCA-MRM | PCA-NN |
|---|---|---|---|
| Quiet | day (daily means) | 3.4 | **1.9** |
| | night (daily means) | 2.0 | **1.4** |
| With flares | day (daily means) | 3.6 | **2.4** |
| | night (daily means) | 2 | **1.6** |
| With geomagnetic storms without flares | day (daily means) | 8.8 | **4.5** |
| | night (daily means) | 2.4 | **2.0** |
| With flares and/or storms | day (daily means) | 4.4 | **2.6** |
| | night (daily means) | 2.1 | **1.6** |

As one can see, the PCA-MRM model performs well during days without significant geomagnetic disturbances, even if a flare is observed. The daily mean and monthly mean MAE depend on the mean values of the geomagnetic indices Dst, Kp, ap: the PCA-MRM model both under- and overestimates TEC values during days with geomagnetic disturbances with approximately similar rates however, large overestimations are seen more often than large underestimations.

Analyzing the "best subsets" of SWp used to build MRM models for different TEC parameters at different time intervals allows us to define space weather parameters the most and the least frequently used in MRMs. The most frequently used space weather parameters (used for ≥ 60% of the days) are Mg II, Dst and B$_y$, p and ap, AE, XR and B$_z$ (see Tab. 4). The least used space weather parameters are the solar wind velocity v, Kp and C.f.

**Table 4.** Most important SWp for the PCA-based models. PCA-MRM – SWp that are most often included in "best subsets". PCA-NN – SWp that allows for the best forecasting skills. SWp that are important for both types of the PCA-based models are in bold. Please note that the F10.7 index was not used for PCA-MRM. Parentheses mark SWp that is important for PCA-NN model without auto-regression term.

| TEC parameter | PCA-MRM (most often used) | PCA-NN (best forecasting skills) |
|---|---|---|
| daily mean TEC | **Mg II** **Dst** XR By | Mg II **Dst** AE ap XR C.f. |
| EOF1 | **Mg II** **Dst** **AE** By | Daily mean TEC (lagged) Mg II **Dst** **AE** EOF1 (lagged) |
| EOF2 | *Mg II* Dst ap Bz p | *F10.7* **Dst** AE (v) EOF2 (lagged) |

*3.4 PCA-NN model*

The PCA-NN model (Fig. 3) uses the same first step as the PCA-MRM model: a 1h TEC series PCA decomposition. Again, only two first modes, which have the highest variance fractions and are responsible for most of the TEC variability, are used for modeling TEC. However, the second step of the PCA-NN model consists of the training of three NNs, one for each of the TEC parameters (the daily mean TEC, EOF1 and EOF2) with SWp as inputs. As previously, L is the length of the input dataset, and SWp series with lags of 1 and 2 days in respect to the TEC series (SWp series lead) are used. Then these three trained NNs are used to reconstruct (forecast) TEC for the following day, day L+1. No negative daily mean TEC and EOF1 series were allowed: in case NNs forecast negative values of daily mean TEC or EOF1, they were multiplied by -1. The forecasted TEC parameters are combined with the corresponding PCs to reconstruct (forecast) the 1h TEC series for the day L+1.

Here we present results aimed at finding a suitable NN configuration and the best set of the SWp predictors that can be used as the NN's input.

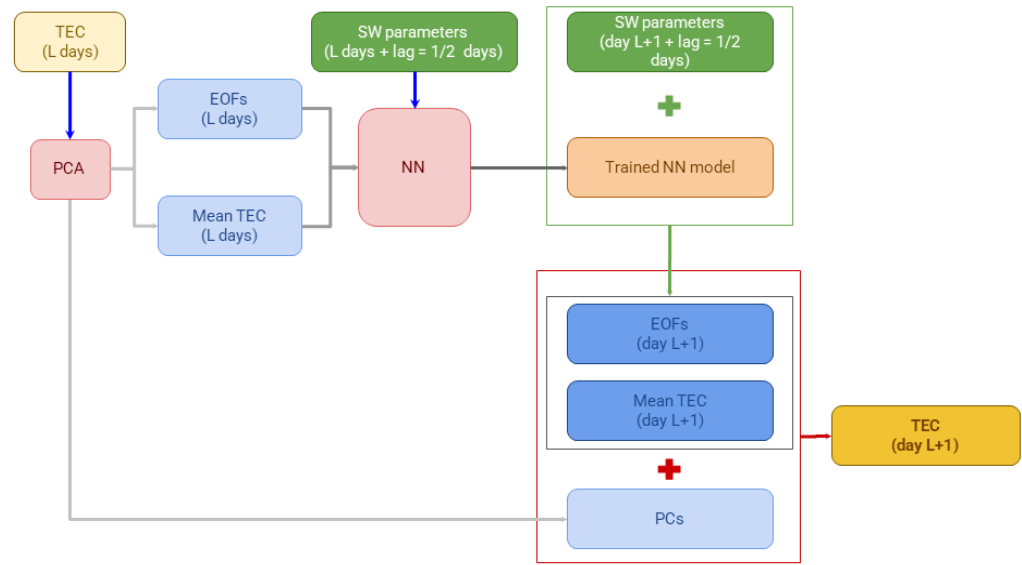

**Figure 3.** The PCA-NN TEC model scheme.

3.4.1 NN algorithm

The PCA-NN model described below is built using a ready-to-use package *neuralnet* (R, https://www.vps.fmvz.usp.br/CRAN/web/packages/neuralnet/, accessed on 26 May 2023) described in [29-30], and examples of its usage can be found, e.g., in [31]. This package allows training feedforward neural networks using a weights backpropagation, resilient backpropagation with or without weight backtracking, or the modified globally convergent version of such network [29-30]. The package allows flexible settings through a custom choice of error and activation functions.

For the PCA-NN model, we used the default error and activation functions (the sum of squared errors ("sse") and the logistic function, respectively) and the "rprop+" algorithm – the resilient backpropagation with weight backtracking, because preliminary tests showed that this algorithm gives better results comparing to other algorithms available in the package.

To find the best NN architecture, we tested:
- Input datasets of different lengths L;
- NNs built with the input SWp series with lag equals only to 1 or 2 days, and NNs built with the input SWp series with lags of 1 and 2 days together;
- NNs of different depths: from 5 to 1 hidden layers;
- NNs with different number of nodes per a hidden layer.

The results of the tests can be summarized as follows:
1. In general, the input dataset with L = 31 days (same as for the PCA-MRM model) results in higher values of the correlation coefficient between the observed and forecasted TEC series and in lower values of MAE and RMSE compared to ones obtained for L equals to 62, 93 or 15 days. This result was obtained even if the L values used for PCA and NN were different: e.g., for PCs calculated for $L_{PCA}$ = 31 days and NNs with $L_{NN}$ = 15, 31, 62, or 93 days. Thus, it seems that is not that only seasonal variations of the TEC daily modes are better resolved on the timescale of about 1 month, as we proposed in the case of the PCA-MRM model [6, 13]. In our mind, the results of the tests of the PCA-NN models with different L show that the relations between SWp and TEC are also slightly changing with time, for example, due to seasonal variability in, e.g., magnetospheric conditions or due to differences in the behaviour of the solar wind-magnetosphere-ionosphere-upper atmosphere system during quiet and disturbed conditions or due to other reason(s). Also, tests made for L between 27 and 35 days show that for different TEC parameters, the optimal L value changes: it is equal to 27 days for the daily mean TEC, to 29 or 31 days for EOF1, and to 32 days for EOF2;

2. It is better to train NN on the series of the SWp predictors with lags of 1 and 2 days together. These NNs allow for the TEC forecast with better skills compared to forecasts made by the averaging of the forecasts made by two separate NNs: one trained on the SWp series with lag = 1 day and another trained on the SWp series with lag = 2 days;
3. NN with a small number of hidden layers perform better: only 2 or 3 hidden layers are sufficient for 3 to 7 predictors (6 to 14 input SWp lagged series, respectively);
4. NNs with different numbers of nodes per hidden layer can provide similar forecast skills, and this parameter can be tuned individually for each TEC parameter and each set of SWp predictors.

A "Monte-Carlo approach" was found to give better results and, therefore, was adopted for this kind of PCA-NN model. This approach consists of training of a number (e.g., 100) of NNs of the same architecture (same number of hidden layers with the same number of nodes per layer) on the same input dataset resulting in 100 "preliminary" forecasts for the day L+1. Then, these "preliminary" forecasts are averaged to obtain the "final" forecast for the day L+1. The standard deviations of the MAE/RMSE of the "preliminary" forecasts are very low: $\sigma \approx$ 2-3% of the MAE/RMSE values both for the daily mean TEC series and for EOFs, i.e., on average, the "preliminary" forecasts are very similar, but the averaging helps to eliminate extremely high |ΔTEC| produced from time to time by "preliminary" forecasts.

3.4.2 Best subsets of predictors

To define sets of best predictors for each TEC parameter (daily mean TEC, EOF1 and EOF2) we trained NNs using just one SWp (two series with different lags) at a time. Comparing the forecasts with the observed series of the TEC parameters and using r (correlation coefficient between the observed and forecasted series), MAE and RMSE metrics, we eliminated those SWp that gave the worst results. Later we tested if the addition of those eliminated parameters to ones found to be the best would nevertheless improve the forecasting quality and found that there was no improvement, and the eliminated SWp is indeed not essential for the PCA-NN models presented here. Those eliminations were made for all three TEC parameters independently.

Afterward, we trained NNs using pairs, triads, tetrads, etc., of the remaining SWp to define the best combination of SWp. Those selections, again, were made independently for each of the three TEC parameters. The lists of SWp that allow training NNs that provide the best forecasting skills are in Tab. 4. Comparing those lists between the TEC parameters and between the PCA-MRM and PCA-NN models, one can see that:

1. Mg II and Dst are SWp that are necessary to build a good PCA-based model (both MRM- and NN-based) for each of the three TEC parameters with one small exception: the PCA-NN model for EOF2 gives much better predictions if the F10.7 index is used instead of Mg II;
2. The AE index is necessary to build a good PCA-NN model for each of the three TEC parameters, while it is important only for the PCA-MRM model of EOF1;
3. A good PCA-NN model for the daily mean TEC needs more SWp predictors than the PCA-NN models for EOFs (6 vs 3-4). Those extra predictors are ap, XR and C.f.;
4. Parameters of the interplanetary magnetic field are not necessary for a good PCA-NN model for any TEC parameter. The only solar wind parameters that may be needed for a good forecast is the solar wind speed and only for EOF2.

However, it can be removed from the set of predictors without losing the forecast quality (see below);
5. None of the solar wind coupling functions found to improve the PCA-NN models forecasts, neither when they substituted parameters of the solar wind/IMF nor added to them.

Statistical analysis of the TEC parameters with 1d time resolution shows that the daily mean TEC and the EOF2 series are auto correlated: the lag1 autocorrelation coefficients $\alpha_1$ are 0.91 for daily mean TEC and 0.38 for EOF2, respectively. On the other hand, for the EOF1 series $\alpha_1$ = 0.14 (no autocorrelation). We tested if the usage of an "auto-regression term" would improve the forecasting quality of NN models and, if yes, for which TEC parameter(s). The auto-regression was introduced by adding to the list of the input series of the two lagged series (with lags of 1 and 2 days) of the TEC parameters predicted by that model. It was found that for the daily mean TEC series and for EOF2, there is an improvement of the forecasting skills (more significant in the case of EOF2). This is in full agreement with the results of the auto-correlation analysis. Even more, it was found that in this case, we can eliminate the solar wind velocity (v) from the set of the input parameters leaving just the solar UV/XR and geomagnetic SWp. For this model, the MAE/RMSE metrics are even better than for a model that includes both the auto-regression and the solar wind velocity.

The tests made only on the data for 2015 showed no improvement of the forecasting quality when an auto-regression term was added to the EOF1 model. However, the results obtained for a longer data set (from December 2014 to June 2018) showed the even for the EOF1 PCA-NN model there is a slight improvement in the forecast quality when the auto-regression term is added. Thus, the PCA-NN models for all the TEC parameters can be built using corresponding auto-regression terms.

The analysis of the best subsets (Tab. 4) shows that using the correlated SWp may improve the forecasting skills of a NN model. For example:
- adding the XR proxy series to the Mg II series (NN model for the daily mean TEC) improves the forecasting skills despite the high correlation between the Mg II and XR series (|rMgII vs XR | = 0.82);
- for all TEC parameters, NN models have better forecasting skills if the series of the Dst and AE indices (|rDst vs AE| = 0.75) are used together;
- using the series of the ap and AE indices (|rap vs AE| = 0.85) together to build the daily mean TEC NN model improves the forecasting skills.

3.4.3 PCA-NN model performance

The final NN configuration, the best subsets of the predictors and forecast quality for the PCA-NN models are shown in Tabs. 2-5 for all TEC parameters. Figure 4 shows the observed and forecasted series of daily mean TEC, EOF1 and EOF2, respectively. All the forecasts are made using a moving window of L = 31 days (taking into account the lags of 1 and 2 days) to perform PCA and to build NN models. Figure 5a shows the final forecast of the 1h TEC series made by using the forecasted TEC parameters and the corresponding PCs.

As one can see from Tabs. 2-3, the skills of PCA-NN models are always better than the ones for PCA-MRM (with the only exception – the 90% confidence level for the night 1h TEC values). Thus, for a PCA-based TEC forecasting model, even a very simple NN-based model (a feedforward neural network) performs better than MRM. For the PCA-NN model, the MAE and RMSE parameters are reduced more than 1.5 times compared to PCA-MRM. The PCA-NN model significantly outperforms the PCA-MRM in forecasting TEC for day hours and slightly outperforms it for night hours. The MaxE parameter (Tab. 2) is lower for the PCA-NN model but still high due to a poorer performance of the PCA-NN model during some geomagnetic storms: as one can see from Fig. 5a, the PCA-NN model systematically underestimate TEC variations during geomagnetic storms. The same conclusion can be made from Tab. 3 – the average MAE for days with storms is 4.5

TECu for the day hours and 2 TECu for the night hours or 1.5-2 times higher than for days without geomagnetic disturbances.

Compared to other TEC models[see 6 and 13 and references therein as well as, e.g., 7-8, 22, 32] the PCA-NN model performs very well with RMSE/MAE in range from 1.2 to 1.7 TECu for the daily mean TEC series and from 1.5 to 3.3 TECu for 1h TEC series, and with the 90% confidence interval (calculated for 2.5 years of the declining phase of the solar cycle 24) of ±4 TECu in average (±4.8 TECu for day hours and ±3.2 TECu for night hours).

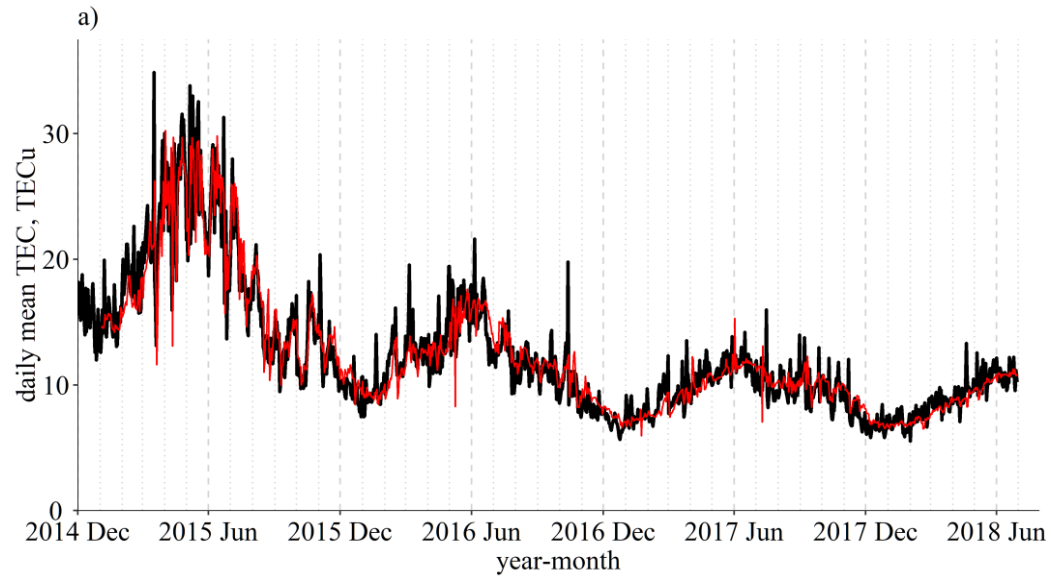

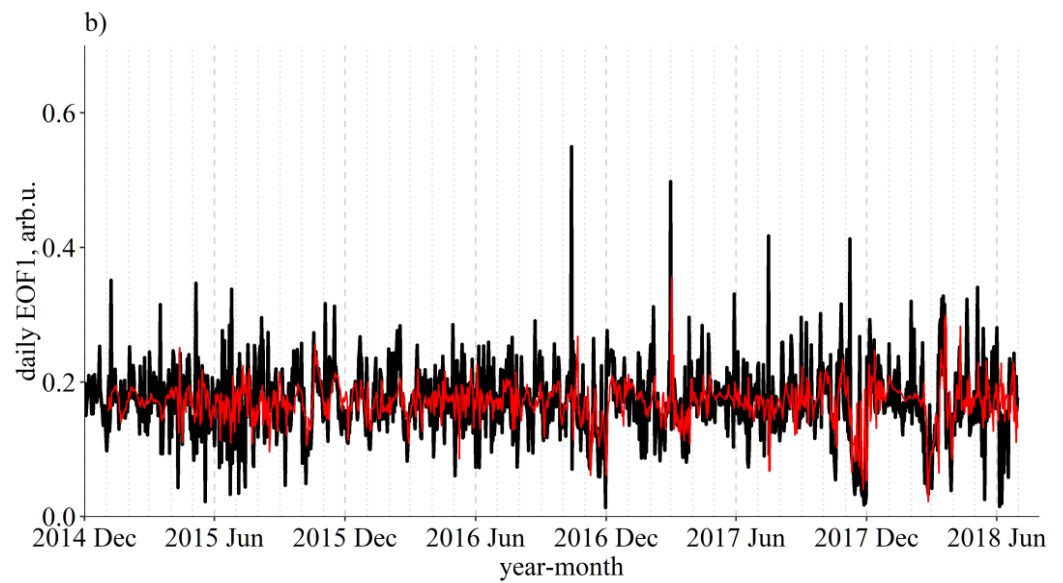

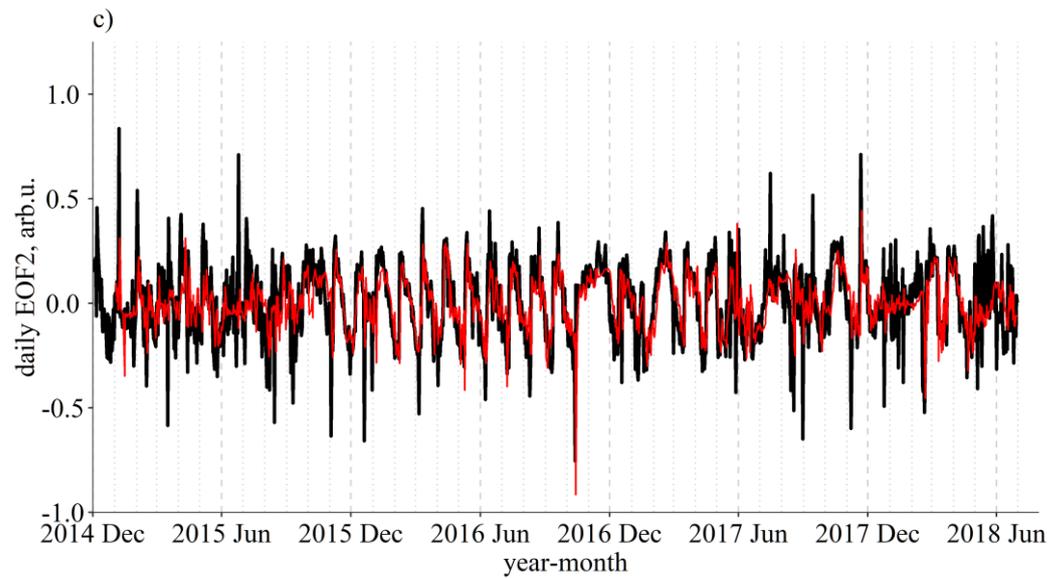

**Figure 4.** The observed (black) and forecasted (red) series of the daily mean TEC (a), EOF1 (b) and EOF2 (c).

**Table 5.** Best NN architecture and best sets of the predictors for the PCA-NN models. The metrics are obtained for the time interval from December 2014 to June 2018. MAE and RMSE are in TECu for daily mean TEC and in arbitrary units for EOF1 and EOF2.

| | daily mean TEC, TECu | EOF1, arb. u. | EOF2, arb. u. |
|---|---|---|---|
| NN (nodes per hidden layer) | (14,6,4) | (8,6,4) | (8,6,4) |
| Number of predictors | 7 predictors | 4 predictors | 4 predictors |
| SWp predictors | MgII, Dst, C.f., ap, AE, XR, daily mean TEC | AE, MgII, Dst, EOF1 | AE, Dst, F10.7, EOF2 |
| r | 0.94 | 0.35 | 0.47 |
| MAE | 1.19 | 0.05 | 0.12 |
| RMSE | 1.72 | 0.07 | 0.16 |

## 4. Validation of the prototype on different datasets

As described above, the prototype was developed on the SCINDA dataset. To validate it, the RENEP dataset (data for three locations shown in Fig. 1) was used. The Cascais location from the RENEP dataset is very close to the position of the SCINDA receiver. Thus it is expected that the performance of the prototype on this series will be very similar to one obtained for the SCINDA series. On the other hand, two other locations, Furnas (Azores) and Funchal (Madeira), are at different longitudes (Furnas, difference in longitude of ~16º) or latitudes and longitudes (Funchal, differences in latitudes of ~6º and longitude of ~7.5º). A recent study [33] of TEC variations during geomagnetic storms at these three locations showed that in most cases, TEC variations at the Lisbon area, Azores and Madeira are very similar, with higher amplitude observed for the most southern location (Madeira). Also, it seems that the Funchal TEC series shows variability that can be attributed to the effect of the equatorial electrojet (EEJ).

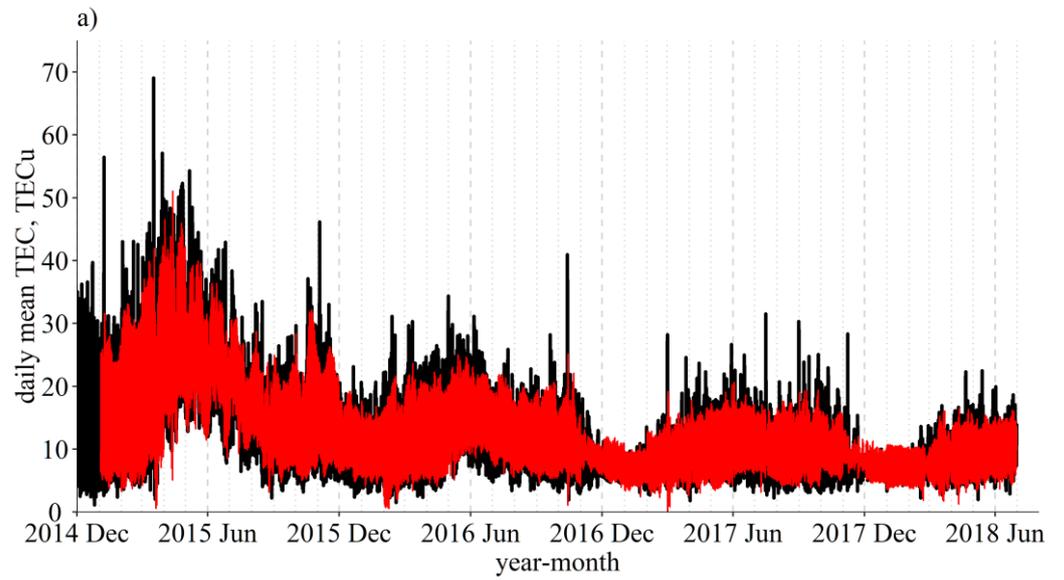

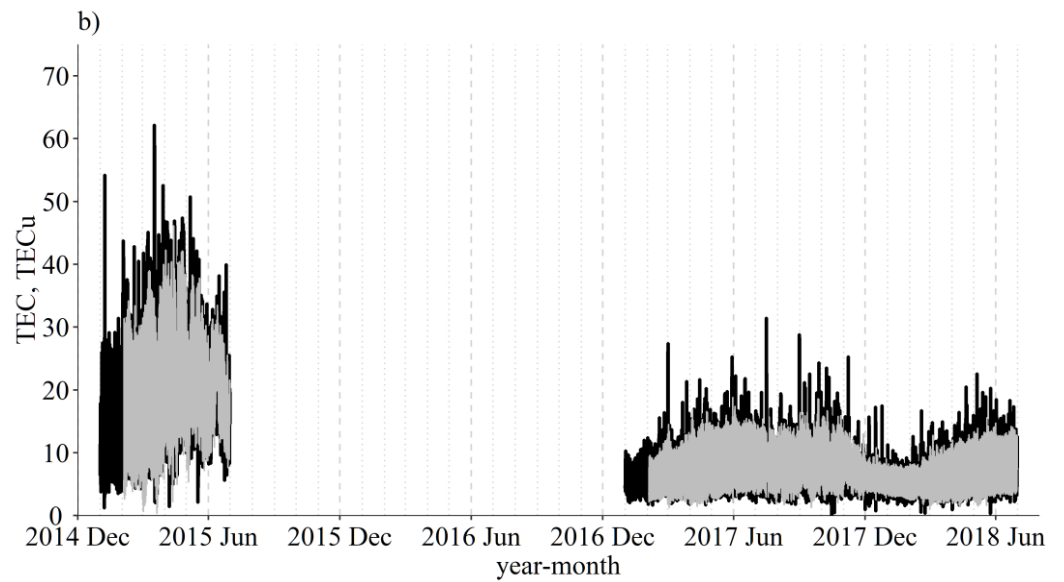

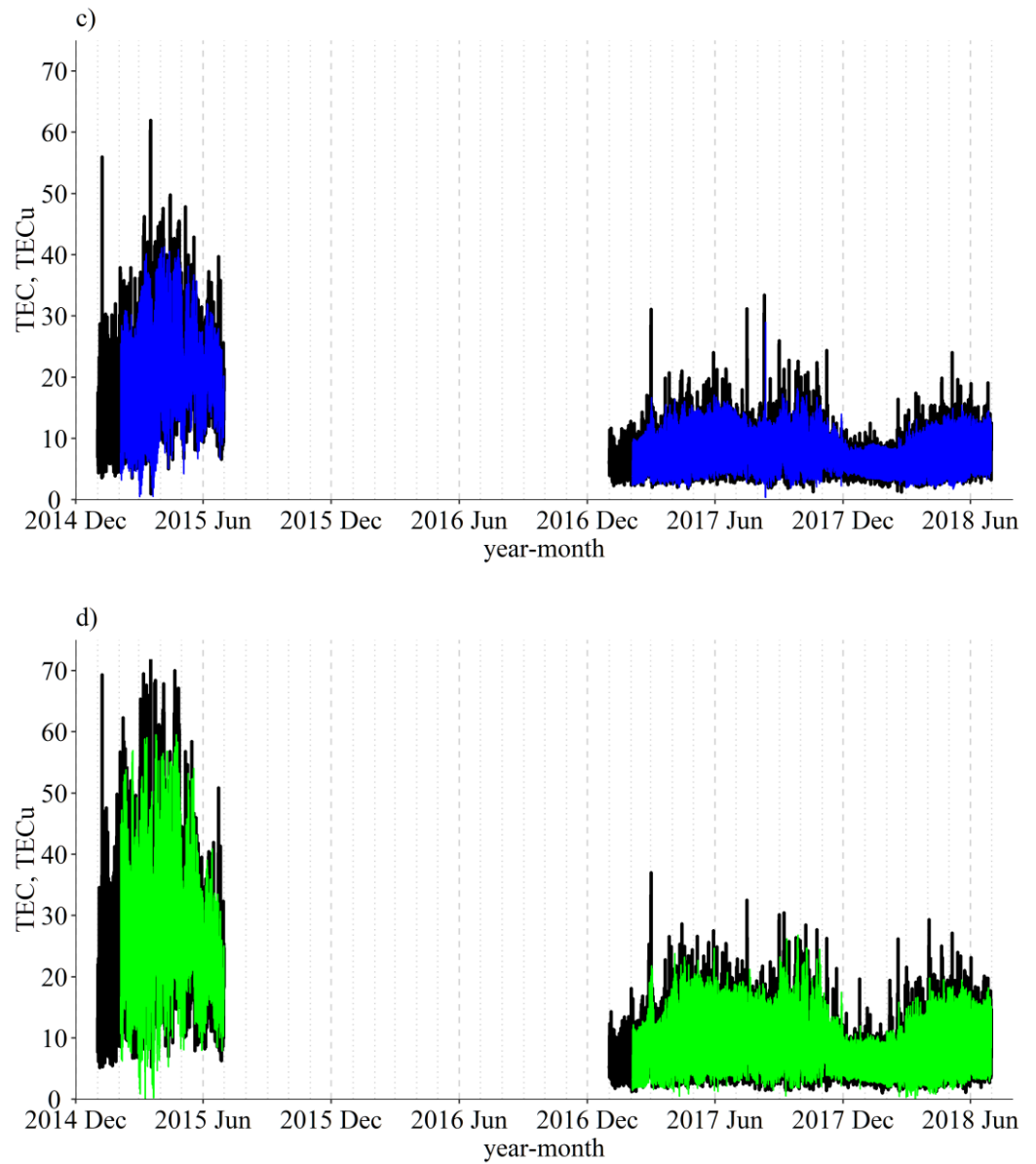

**Figure 5.** The observed (black) and forecasted (colored) series of 1h TEC series for the SCINDA dataset (a, red) - prototype, and the RENEP dataset (Cascais (b, grey), Furnas (c, blue) and Funchal (d, green) - validation.

To validate the PCA-NN prototype we submitted, separately, the Cascais, Furnas and Funchal TEC series to the PCA-NN model using the same scheme (Fig. 3), the same NN architecture as for the SCINDA series, and the same respective lists of input parameters to forecast the daily mean TEC, EOF1 and EOF2 for these three locations. The results (1h TEC series) are shown in Figs. 5b-d for Cascais, Furnas and Funchal, respectively. Please note that because of the gap in the RENEP data, the models were applied separately for the January to June 2015 and for January 2017 to June 2018 time intervals.

As one can see from Fig. 5, in general, the performance of the PCA-NN models for the Lisbon airport (SCINDA-based prototype), Cascais and Furnas are very similar, while for Funchal the model shows higher variability and bigger errors (even considering overall higher TEC variability of the Funchal series – black line in Fig. 5d). This is confirmed by the distribution of the forecasting errors (ΔTEC) shown in Fig. 6 for all four series. Table 6 allows to compare the forecasting qualities of all four models (in this case, the SCINDA prototype model metrics were calculated for the same time intervals as for the validation models with RENEP series).

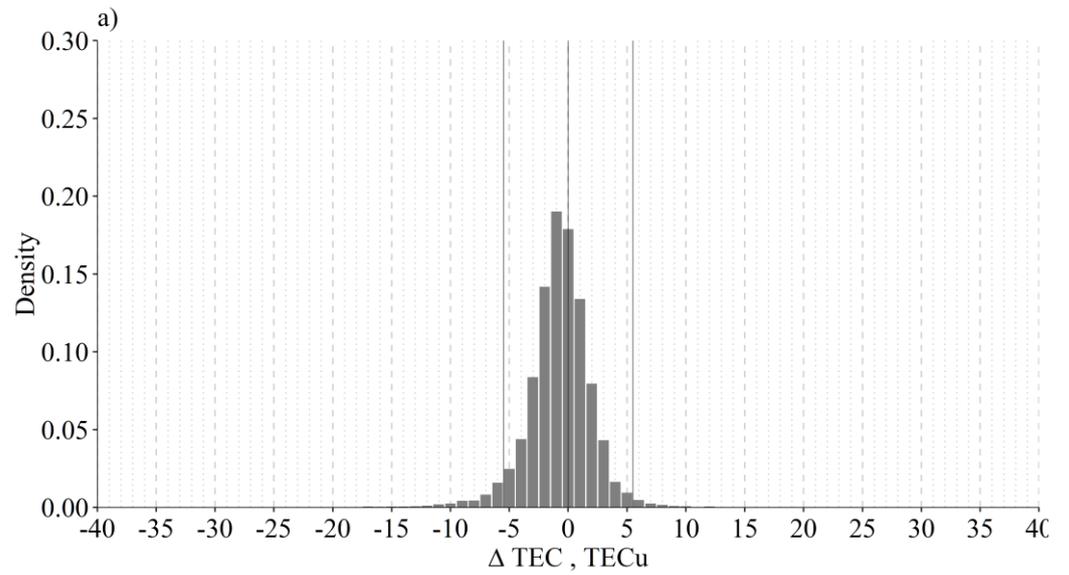

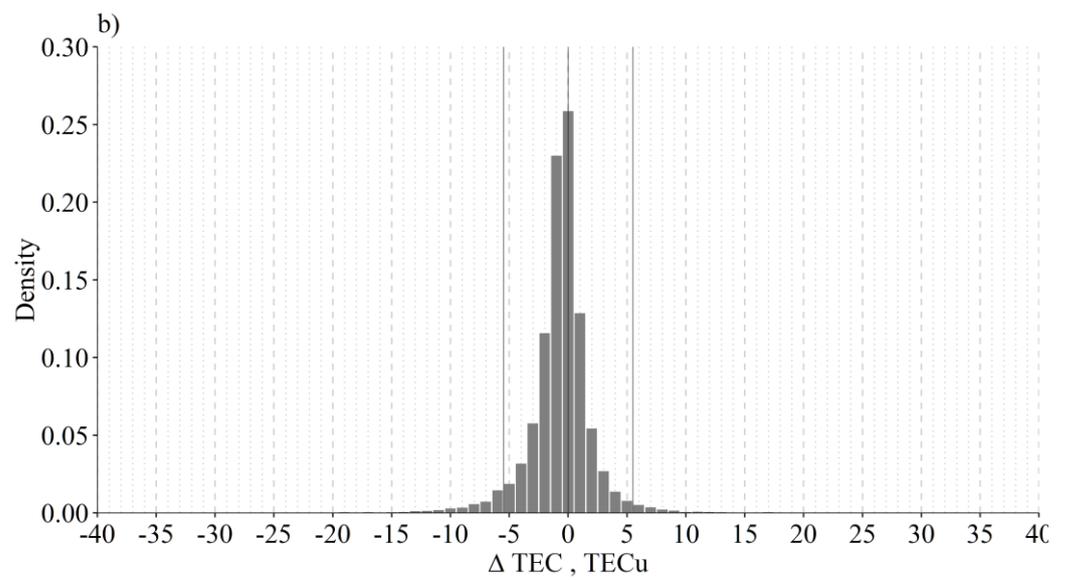

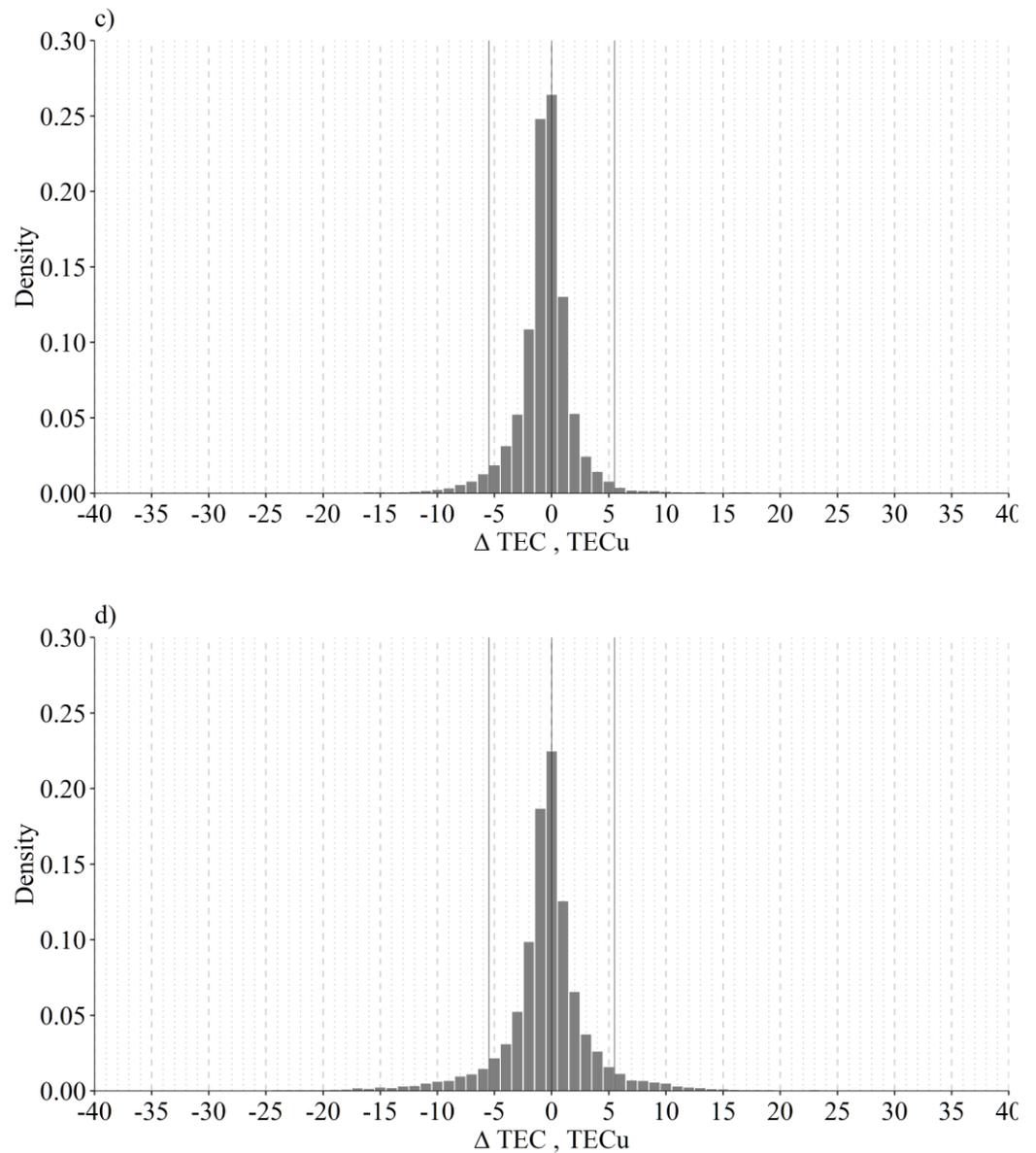

**Figure 6.** Distribution of the model errors (ΔTEC) as the probability density for the TEC series of SCINDA (a) - prototype, and Cascais (b), Furnas (c) and Funchal (d) - validation.

**Table 6.** Performance of the PCA-NN model for four locations: SCINDA Lisbon (prototype), and Cascais, Furnas and Funchal (validation) using data for January - June 2015 and January 2018 - June 2018.

| Metrics | SCINDA (Lisbon) | Cascais | Furnas (Azores) | Funchal (Madeira) |
|---|---|---|---|---|
| r(TEC$_{for}$ vs TEC$_{obs}$), 2015 & 2017-2018 | 0.93 | 0.94 | 0.94 | 0.93 |
| r(TEC$_{for}$ vs TEC$_{obs}$), 2015 | 0.90 | 0.88 | 0.88 | 0.83 |
| r(TEC$_{for}$ vs TEC$_{obs}$), 2017-2018 | 0.80 | 0.88 | 0.88 | 0.89 |
| 95% confidence level, TECu | ±5.0 | ±5.5 | ±5.0 | ±8.5 |
| 90% confidence level, TECu | ±4.0 | ±4.0 | ±3.5 | ±6.0 |

As one can see, while the correlation coefficients for the 1h TEC series are high and similar for all series, there is a significant difference in the ΔTEC distribution. The confidence intervals are similar only for the SCINDA, Cascais and Furnas series (about ±3.5 -

±5.5 TECu), while for Funchal, the most southern station, the confidence intervals are ±6.0 - ±8.5 (TECu) or about 1.5 times higher. This also can be deduced from the width of histograms shown in Fig. 6. We interpret this difference as an influence (unaccounted by used input SWp) of EEJ and events related to the equatorial ionosphere (as, for example, equatorial plasma bubbles that in some regions and under certain conditions can reach Madeira longitude and latitude [33-34]).

## 5. Conclusions

A prototype for a PCA-NN model to forecast the total electron content (TEC) at a single location using space weather parameters (SWp) as predictors is presented. The model consists of 2 steps:
1. A series of TEC of the length of 31 days with 1h time resolution is decomposed by the principal component analysis (PCA) into daily modes;
2. NN models are built to forecast the daily mean TEC and the amplitudes of the first two PCA modes using the lagged SWp. These NN models are used to forecast the daily mean TEC and the amplitudes of the first two PCA modes and, as a result the 1h TEC series, for the following day.

In the presented prototype, a simple feedforward neural network algorithm is used. It was found that the NN models needed to forecast TEC with good scores only need a few hidden layers (3 layers would suffice) with a limited number of nodes per layer.

Compared to the previous PCA-MRM model [6], the PCA-NN models allow for a significant reduce the number of the input SWp, limiting them to some geomagnetic indices (Dst and AE for all TEC parameters plus ap for the daily mean TEC) and to the proxies of the solar UV and XR flux (Mg II/F10.7 for all TEC parameters plus XR and the number of flares of up to C class for the daily mean TEC). For all TEC parameters, the NN models with the same lagged TEC parameter as an input perform better than those without such an "auto-regression" parameter.

It was shown that the PCA-NN model outperforms the PCA-MRM model and the forecasting skills of the PCA-NN model are better than the ones for the PCA-MRM model: MAE/RMSE are decreased by 1.5-2 times, and the correlation coefficients between the observed and forecasted TEC (both for the 1d and 1h time resolution series) as well as the explained variance values are higher.

Contrary to the PCA-MRM, the PCA-NN model systematically underestimates TEC variations during days with geomagnetic disturbance, both for the 1d and 1h time resolution series, though the average MAE of the PCA-NN model for such days is almost twice lower (4.5 TECu vs 8.8 TECu).

Thus, changing the method of forecasting TEC parameters from the linear regression to neural networks allows for improving the forecasting skills of a PCA-based model, decreasing errors (difference between the observed and forecasted TEC values) by 1.5-2 times.

The prototype model was validated on different TEC datasets, obtained from different sources and locations. The validation tests show that for locations at about the latitude of 40ºN the model's performance is similar to one shown by the prototype. Supposedly, this still will be true for the European longitude sector and latitudes between 40ºN and 50ºN. For locations to the south of this zone (as, for example, Madeira) the model's performance deteriorates, most probably, due to the effect of structures in the equatorial ionosphere (EEJ and plasma bubbles that in rare cases, can affect ionosphere at Canarias and Madeira archipelagos, and even southern coast of the Iberian Peninsula). Most probably, the model's performance for southern regions can be improved by adding parameters describing EEJ strength or bubbles' activity. The model is not yet tested for TEC series obtained at latitude north to 40ºN, but, most probably, for a more northern location the list of the input SWp is also needed to be updated to include parameters describing polar ionosphere variability (other than AE, AL or AU indices).

Finally, the list of the input SWp parameters that are strictly necessary for a good performance of the PCA-NN model includes the Dst and AE geomagnetic indices and the F10.7/Mg II solar UV indices. This allows one to conclude that the main space weather drivers of the ionospheric variability in the studied zone (the European longitude sector at latitudes between 30ºN and 40ºN) are geomagnetic field variations associated with geomagnetic storms (the Dst index is a good proxy for this part of variability) and auroral electrojet (the AE index is a good proxy for this part of variability, and changes of the solar UV flux. Also, the F10.7 (the best UV flux proxy for the daily mean TEC and for EOF1) and Mg II (the best UV proxy for EOF2) indices are proxies for similar but not identical processes of the UV flux generation at the Sun surface.

**Supplementary Materials:** The following supporting information can be downloaded at: www.mdpi.com/xxx/s1, Table S1: Absolute values of the correlation coefficients between space weather parameters.

**Author Contributions:** Conceptualization, A.M.; methodology, A.M., Tatiana Barlyaeva; software, A.M.; validation, A.M., Tatiana Barlyaeva, Teresa Barata and R.G.; resources, A.M.; data curation, A.M., Tatiana Barlyaeva; writing—original draft preparation, A.M.; writing—review and editing, A.M., Tatiana Barlyaeva, Teresa Barata and R.G.; visualization, A.M.; project administration, A.M., Teresa Barata; funding acquisition, A.M., Teresa Barata. All authors have read and agreed to the published version of the manuscript.

**Funding:** IA is supported by Fundação para a Ciência e a Tecnologia (FCT, Portugal) through the research grants UIDB/04434/2020 and UIDP/04434/2020. This study is a contribution to the PRIME project (EXPL/CTA-MET/0677/2021, FCT, Portugal).

**Institutional Review Board Statement:** Not applicable.

**Informed Consent Statement:** Not applicable.

**Data Availability Statement:** The SCINDA TEC data for 2015 are available at [11-12]. The datasets are described in [9-10], respectively. The RENEP RINEX 2.11 files are available through https://renep.dgterritorio.gov.pt/ (accessed on 26 May 2023).

We acknowledge the mission scientists and principal investigators who provided the data used in this research:

The TEC data sets are from the Royal Observatory of Belgium (ROB) data base and are publicly available in IONEX format at ftp:/gnss.oma.be/gnss/products/IONEX/, see also [35] for more in-formation.

The Dst index is from the Kyoto World Data Center http:/wdc.kugi.kyoto-u.ac.jp/dst_final/index.html.

Geomagnetic data measured by the Coimbra Geomagnetic Observatory are available at the World Data Centre for Geomagnetism web portal http:/www.wdc.bgs.ac.uk/dataportal/ .

The solar wind data and the ap index are from the SPDF OMNIWeb database. The OMNI data were obtained from the GSFC/SPDF OMNIWeb interface at https:/omniweb.gsfc.nasa.gov, see also [36] for more details.

The Mg II data are from Institute of Environmental Physics, University of Bremen http:/www.iup.uni-bremen.de/gome/gomemgii.html, see also [18] for more information.

The data on the variations of the solar XR flux are from the LASP Interactive Solar Irradiance Da-ta Center (LISRD, http:/lasp.colorado.edu/lisird/ ). LISIRD provides a uniform access interface to a comprehensive set of Solar Spectral Irradiance (SSI) measurements and models from the soft X-ray (XUV) up to the near infrared (NIR), as well as Total Solar Irradiance (TSI). The XRTIMED data are from the Solar EUV Experiment (SEE) measures the solar ultraviolet full-disk irradiance for the NASA TIMED mission. Level 3 data represent daily averages and are filtered to remove flares available at http:/lasp.colorado.edu/lisird/data/timed_see_ssi_l3/.

The X-ray Flare dataset was prepared by and made available through the NOAA National Geophysical Data Center (NGDC). The data about the solar flares for 2015 are from

> https://www.ngdc.noaa.gov/stp/space-weather/solar-data/solar-features/solar-flares/x-rays/goes/xrs/goes-xrs-report_2015_modifiedreplacedmissingrows.txt.
>
> **Acknowledgments:** The authors would like to acknowledge the Direção Geral do Território (DGT) for making ReNEP data available, as well as the people involved (Helena Ribeiro). The authors are grateful to Yuri Yasyukevich and his team for the development of the GNSS Lab software and technical support.
>
> **Conflicts of Interest:** The authors declare no conflict of interest.

**References**

1. Kumar, V.V., Parkinson, M.L. A global scale picture of ionospheric peak electron density changes during geomagnetic storms. *Space Weather* **2017**, 15 (4), 637–652.
2. Tang. R., Zeng, F., Chen, Z., Wang, J.S., Huang, C.M., Wu, Z. The comparison of predicting storm-time ionospheric TEC by three methods: ARIMA, LSTM, and Seq2Seq. *Atmos.* **2020**, 11(4), 316.
3. Xiong, P., Zhai, D., Long, C., Zhou, H., Zhang, X., Shen, X. Long short-term memory neural network for ionospheric total electron content forecasting over China. *Space Weather* **2021**, 19(4), e2020SW002706.
4. Sivavaraprasad, G., Mallika, I.L., Sivakrishna, K., Ratnam, D.V. A novel hybrid Machine learning model to forecast ionospheric TEC over Low-latitude GNSS stations. *Adv. Space Res.* **2022**, 69(3), 1366-1379.
5. Nath, S., Chetia, B., Kalita, S. Ionospheric TEC prediction using hybrid method based on ensemble empirical mode decomposition (EEMD) and long short-term memory (LSTM) deep learning model over India. *Adv. Space Res.* **2023**, 1 71(5), 2307-2317.
6. Morozova, A.L., Barata, T., Barlyaeva, T. PCA-MRM model to forecast TEC at middle latitudes, *Atmos.*, **2022**, 13(2), 323, doi: 10.3390/atmos13020323
7. Iluore, K., Lu, J. Long short-term memory and gated recurrent neural networks to predict the ionospheric vertical total electron content. *Adv. Space Res.* **2022**, 70, 652–665.
8. Ren, X., Yang, P., Liu, H., Chen, J., Liu, W. Deep learning for global ionospheric TEC forecasting: Different approaches and validation. *Space Weather* **2022**, 20, e2021SW003011. doi: 10.1029/2021SW003011
9. Barlyaeva, T., Barata, T., Morozova, A. Datasets of ionospheric parameters provided by SCINDA GNSS receiver from Lisbon airport area. *Data in Brief* **2020**, 31, 105966, doi: 10.1016/j.dib.2020.105966
10. Morozova, A., Barlyaeva, T., Barata, T. Datasets of ionospheric parameters (TEC, SI, positioning errors) from Lisbon airport area for 2014-2019, *Data in Brief* **2023**, doi: 10.1016/j.dib.2023.109026
11. Data from: Barlyaeva, T., Barata, T., Morozova, A. Datasets of ionospheric parameters provided by SCINDA GNSS receiver from Lisbon airport area, Mendeley Data **2020**. http:/dx.doi: 10.17632/kkytn5d8yc.1
12. Data from: Morozova, A., Barlyaeva, T., Barata, T. Datasets of ionospheric parameters (TEC, SI, positioning errors) from Lisbon airport area for 2014-2019, Mendeley Data, **2022** http://dx.doi: 10.17632/3z6mjk39jv.2
13. Morozova, A.L., Barlyaeva, T.V., Barata, T. Variations of TEC over Iberian Peninsula in 2015 due to geomagnetic storms and solar flares. *Space Weather* **2020**, 18(11), p.e2020SW002516
14. Yasyukevich, Y.V., Mylnikova, A.A., Kunitsyn, V.E., Padokhin, A.M. Influence of GPS/GLONASS differential code biases on the determination accuracy of the absolute total electron content in the ionosphere. Geomagn. Aeron. 2015, 55, 763–769. https://doi.org/10.1134/S001679321506016X.
15. Mylnikova, A.A., Yasyukevich, Y.V., Kunitsyn, V.E., Padokhin, A.M. Variability of GPS/GLONASS differential code biases. Results Phys. 2015, 5, 9–10. https://doi.org/10.1016/j.rinp.2014.11.002.
16. Yasyukevich, Y.V., Mylnikova, A.A., Polyakova, A.S. Estimating the total electron content absolute value from the GPS/GLONASS data. *Results Phys.* **2015**, 5, 32–33. https://doi.org/10.1016/j.rinp.2014.12.006.
17. Morozova, A.L., Ribeiro, P., Blanco, J.J., Barlyaeva, T.V. Temperature and pressure variability in mid-latitude low atmosphere and stratosphere-ionosphere coupling, *Adv. Space Res.* **2020**, 65, 9, 2184-2202, doi: 10.1016/j.asr.2019.10.039
18. Viereck, R., Puga, L., McMullin, D., Judge, D., Weber, M., Tobiska, W.K. The Mg II index: A proxy for solar EUV. *Geophys. Re. Let.* **2001**, 28(7), pp.1343-1346.
19. Snow, M., Weber, M., Machol, J., Viereck, R., Richard, E. Comparison of Magnesium II core-to-wing ratio observations during solar minimum 23/24, *J. Space Weather Space Clim.* **2014**, 4, A04, doi:10.1051/swsc/2014001
20. Newell, P. T., T. Sotirelis, K. Liou, C.-I. Meng, and F. J. Rich A nearly universal solar wind-magnetosphere coupling function inferred from 10 magnetospheric state variables, *J. Geophys. Res.* **2007**, 112, A01206, doi:10.1029/2006JA012015
21. Verkhoglyadova, O.P., Komjathy, A., Mannucci, A.J., Mlynczak, M.G., Hunt, L.A., Paxton, L.J., Revisiting ionosphere-thermosphere responses to solar wind driving in superstorms of November 2003 and 2004. *J. Geophys. Res.: Space Phys.* **2017**, 122(10), pp.10-824
22. Maruyama, T. Solar proxies pertaining to empirical ionospheric total electron content model, *J. Geophys. Res.* **2010**, 115, doi:10.1029/2009JA014890.
23. Mukhtarov, P., Andonov, B., Pancheva, D. Empirical model of TEC response to geomagnetic and solar forcing over Balkan Peninsula. *J. Atmos. Solar-Terrestrial Phys.* **2018**, 167, 80-95

21 of 23

24. Bjornsson, H., Venegas, S. A. *A manual for EOF and SVD analyses of climatic data*, McGill University, CCGCR Report 97-1, 1997.
25. Hannachi, A., Jolliffe, I.T., Stephenson, D.B. Empirical orthogonal functions and related techniques in atmospheric science: A review, *Int. J. Climatol.* **2007**, 27 (9), 1119-1152.
26. Shlens, J. A tutorial on principal component analysis. arXiv preprint arXiv:1404.1100, https:/arxiv: pdf/1404.1100, 2014.
27. Ebisuzaki, W. A method to estimate the statistical significance of a correlation when the data are serially correlated, *J. Clim.* **1997**, 10 (9), 2147-2153.
28. Sheskin, D.J. Handbook of parametric and nonparametric statistical procedures. Chapman and Hall/CRC, 2003.
29. Günther, F., Fritsch, S. Neuralnet: training of neural networks. *The R Journal* **2010**, 2(1), p.30.
30. Fritsch, S., Guenther, F., Guenther, M.F. Package 'neuralnet'. Training of Neural Networks, preprint available at https://cran.r-project.org/web/packages/neuralnet/neuralnet.pdf, 2019.
31. Ciaburro, G., Venkateswaran, B. Neural Networks with R: Smart models using CNN, RNN, deep learning, and artificial intelligence principles. Packt Publishing Ltd., 2017.
32. Mukhtarov, P., Andonov, B., Pancheva, D. Empirical model of TEC response to geomagnetic and solar forcing over Balkan Peninsula. *J. Atmos. Solar-Terrestrial Phys.* **2018**, 167, 80-95
33. Barata, T., Pereira, J., Hernández-Pajares, M., Barlyaeva, T., Morozova, A. Ionosphere over Eastern North Atlantic Midlatitudinal Zone during Geomagnetic Storms. *Atmos.* **2023**, 14, 949. https://doi.org/10.3390/atmos14060949
34. Cherniak, I., Zakharenkova, I. First observations of super plasma bubbles in Europe. *Geophys. Res. Lett.* **2016**, 43, 11137–11145. https://doi.org/10.1002/2016GL071421.
35. Bergeot N., Chevalier, J.-M., Bruyninx, C., Pottiaux, E., Aerts, W., Baire, Q., Legrand, J., Defraigne P., Huang, W. Near real-time ionospheric monitoring over Europe at the Royal Observatory of Belgium using GNSS data, *J. Space Weather Space Clim.* **2014**, 4, A31, doi: 10.1051/swsc/2014028
36. Papitashvili, N., King, J.H. May. Solar Wind Spatial Scales in, and Comparisons of, Hourly Wind and ACE IMF and Plasma Data. In: AGU Spring Meeting Abstracts, vol. 2004, pp. SH41A-07.

# Supplementary Materials

for

# Total electron content PCA-NN model for middle latitudes

by

**Anna Morozova** [1,2,*]**, Teresa Barata** [1,3]**, Tatiana Barlyaeva** [1] **and Ricardo Gafeira** [1,2]

This file contains Supplementary Materials (SM) referenced in the main text:

**Supplementary Table S1.** Absolute values of the correlation coefficients |r| between SWp. Only |r| ≥ 0.68 with p values < 0.01 are shown. Correlation coefficients of SWp of the same type (IMF; solar wind; geomagnetic indices; solar UV and XR fluxes, and solar flares) have solid borders. Correlation coefficients between the coupling functions and other SWp have dashed borders.

|  | By | Bz | p | Dst | ap | AE | Kp | K$_{coi}$ | F10.7 | XR | C.f. |
|---|---|---|---|---|---|---|---|---|---|---|---|
| **Bx** | 0.69 | | | | | | | | | | |
| **n** | | | 0.75 | | | | | | | | |
| **ap** | | | | 0.74 | | 0.85 | 0.89 | 0.86 | | | |
| **AE** | | | | 0.75 | 0.85 | | | 0.86 | | | |
| **Kp** | | | | 0.73 | 0.89 | 0.91 | | 0.97 | | | |
| **K$_{coi}$** | | | | 0.68 | 0.86 | | | | | | |
| **dΦ$_{MP}$/dt** | | 0.83 | | | 0.75 | 0.87 | 0.75 | 0.69 | | | |
| **E$_{WAV}$** | | 0.85 | | | 0.72 | 0.79 | | | | | |
| **E$_{WV}$** | | 0.82 | | | 0.78 | 0.80 | 0.7 | | | | |
| **ε$_3$** | | 0.8 | | | 0.79 | 0.80 | 0.72 | | | | |
| **E$_{KLV}$** | | 0.74 | | | 0.79 | 0.79 | 0.74 | 0.69 | | | |
| **E$_{KL}$** | | 0.77 | | | 0.72 | 0.77 | 0.69 | | | | |
| **v·B$_S$** | | 0.85 | | | 0.7 | 0.76 | | | | | |
| **E$_{SR}$** | | 0.76 | | | 0.77 | 0.72 | | | | | |
| **E$_{TL}$** | | 0.75 | | | 0.73 | 0.68 | | | | | |
| **Mg II** | | | | | | | | | 0.79 | 0.82 | |
| **F10.7** | | | | | | | | | | 0.84 | |
| **N.f.** | | | | | | | | | | | 0.98 |